\documentclass[preprint2]{aastex}

\shorttitle{A large anisotropy in the 3CRR quasar distribution}
\shortauthors{Singal}

\begin{document}

\title{A large anisotropy in the sky distribution of  3CRR quasars and other radio galaxies}
\author{Ashok K. Singal}
\affil{Astronomy and Astrophysics Division, Physical Research Laboratory,\\
Navrangpura, Ahmedabad - 380 009, India}
\email{asingal@prl.res.in}
\begin{abstract}
We report the presence of large anisotropies in the sky distributions of powerful extended quasars as well as 
some other sub-classes of radio galaxies in the 3CRR survey, the most reliable and most intensively studied complete  
sample of strong steep-spectrum radio sources. The anisotropies lie about a plane passing through the equinoxes 
and the north celestial pole. Out of a total of 48 quasars in the sample, 33 of them lie in one half of the observed sky and 
the remaining 15 in  the other half. The probability that in a random distribution of 3CRR quasars in the sky, statistical 
fluctuations could give rise to an asymmetry in observed numbers up to this level is only $\sim 1\%$. Also only about 1/4th of 
Fanaroff-Riley 1 (FR1) type of radio galaxies lie in the first half of the observed sky and the remainder 
in the second half. If we include all the observed 
asymmetries in the sky distributions of quasars and radio galaxies in the 3CRR sample, the probability of their 
occurrence by a chance combination reduces to $\sim 2 \times 10^{-5}$. 
Two pertinent but disturbing questions that could be raised here are --  firstly  why should there be such large 
anisotropies present in the sky distribution of some of the strongest and most distant 
discrete sources, implying inhomogeneities in the universe at very large scales (covering a fraction of the universe)? 
Secondly why should such anisotropies lie about a great circle 
decided purely by the orientation of earth's rotation axis and/or the axis of its revolution around the sun? 
Are these alignments a mere coincidence or do they imply that these axes 
have a preferential placement in the larger scheme of things, implying an apparent breakdown of the Copernican principle 
or its more generalization, cosmological principle, upon which the standard cosmological model is based upon?
\end{abstract}
\section{Introduction}
Copernican principle states that earth does not have any eminent or privileged position in the universe and therefore an observer's 
choice of origin and/or orientation of his/her coordinate system should have no bearing on the appearance of the distant universe. 
Its natural generalization is the cosmological principle which states that the universe on a sufficiently large scale should appear 
homogeneous and isotropic, with no preferred directions, to all observers.  
However to us on earth the universe does show heterogeneous structures up to the scale of superclusters of galaxies and even 
somewhat beyond, but the conventional wisdom is that it would all appear homogeneous and isotropic when observed on 
still larger scales, perhaps beyond a couple of hundreds of megaparsecs. 
Radio galaxies and quasars, the most distant discrete objects (at distances of gigaparsecs and farther) seen in the universe,  
should trace the distribution of matter in the universe at that large scale and should therefore appear isotropically distributed 
from any vantage point in the universe, including that on earth. 

On the other hand Cosmic Microwave Background Radiation (CMBR) observations from the WMAP satellite have in recent years been 
reported to show some unexpected anisotropies, which surprisingly seem to be aligned with the ecliptic (Tegmark et al. 
2003; de Oliveira-Costa et al. 2004; Ralston \& Jain 2004; Schwarz et al. 2004; Land \& Magueijo 2005). The alignment of 
the four normals to the quadrupole and octopole planes in the CMBR with the cosmological dipole and the equinoxes  
(Copi et al. 2010) could undermine our ideas about the standard cosmological model with very damaging implications. 
The latest data from the Planck satellite have confirmed the presence of these anisotropies (Ade et al. 2014). 
Also using a large sample of radio sources from the NRAO VLA Sky Survey (NVSS, Condon et al. 1998), which covers whole sky north of 
declination $-40^{\circ}$ and contains 1.8 million sources with a flux-density limit $S>3$ mJy at 1.4 GHz, Singal (2011) 
showed in this faint radio source distribution the presence of a dipole anisotropy which is about 4 times larger 
than the CMBR dipole (Lineweaver et al. 1996; Hinshaw et al. 2009), presumably of a kinetic origin due to the solar motion with respect 
to the otherwise isotropic CMBR. These unexpected findings have recently been corroborated by two independent groups (Rubart \& Schwarz
2013; Tiwari et al. 2015; also see Singal 2014a for a clarification on some misgivings in the literature about  
the formulation used in these analyses).  The fact that the direction of the two independently derived dipoles, viz. from NVSS and 
CMBR, coincide implies that there is certainly some peculiarity along this direction in sky which incidentally lies 
close to the line joining the equinoxes.  
But the large difference in the inferred motion (as much as a factor of  $\sim 4$) cannot be easily explained.  
A genuine discrepancy in the dipoles inferred with respect to two different cosmic reference frames  
would imply a relative motion between these frames, not in accordance with our present ideas of cosmology.

A large--scale bulk flow has also been inferred from peculiar velocities of clusters of galaxies 
(Kashlinsky et al. 2011), though the genuineness of these results has been severely criticized in the literature (Keisler 2009; 
Osborne et al. 2011).  There are reports of the presence of other large--scale alignments in radio and 
optical polarizations data (Jain and \& Ralston 1999; Hutsemekers et al. 2005). 
It seems the universe might not be all isotropic and homogeneous, as assumed in the 
cosmological principle. Here we report even larger anisotropies which are seen in the sky distributions of powerful 
extended quasars and some other sub-classes of radio galaxies.

\section{The Sample}
One of the earliest and best studied source of radio galaxies and quasars 
is the third Cambridge twice revised (3CRR) catalogue (Laing et al. 1983), which is radio complete 
in the sense that all radio 
sources brighter than the sensitivity limit ($S_{178}=10.9$ Jy) of the survey are included 
(and certainly with no spurious entries as each and every source in the sample 
has been studied in detail). It covers the sky north 
of declination, $\delta=10^\circ$, except for a zone of avoidance, a band of $\pm 10^\circ$ about the galactic plane ($b=0^\circ$). 
Also it has a 100\% optical identification content with detailed optical spectra to classify 
radio sources into radio galaxies and quasars. The catalogue with the latest updates is downloadable 
from http://astroherzberg.org/people/chris-willott/research/3crr/.   

The steep spectrum radio sources (radio spectral index $\alpha > 0.5$ with $S\propto \nu^{-\alpha}$) in the 3CRR catalogue 
are divided broadly into two classes, radio galaxies and quasars, the former further 
sub-divided into two types, Fanaroff-Riley 1 and 2 (FR1 and FR2), based on their radio morphologies (Fanaroff \& Riley 1974) with the 
quasars almost always resembling the radio morphology of FR2 types. When compared to FR1s, the FR2 types are almost always found amongst 
the more powerful radio galaxies, overlapping the radio luminosities of quasars.
However FR2 radio galaxies in general show narrow emission lines in their optical spectra, while quasars always show broad emission lines. 
Included among quasars is a small number of what are termed as broad line radio galaxies (BLRGs) or weak quasars (WQ), the latter 
with broad emission lines seen in polarized optical emission, or/and compact optical nuclei detected in infrared or X-rays. 
FR2 radio galaxies are further sub-divided by their optical spectra 
into low excitation galaxies (LEGs) and high excitation galaxies (HEGs). One object (3C386) 
shows an overlap of LEG and WQ properties (see Grimes et al. 2004), which we have therefore dropped from our sample.
Also excluded are a small number of compact steep spectrum sources (CSSS, with angular size $\stackrel{<}{_{\sim}}2$ 
arcsec) which seem to be a different class (Kapahi et al. 1995). Then we have  
23 FR1s, 17 LEGs, 65 HEGs and 48 quasars, making a total of 153 radio sources in our sample.

The conventional wisdom (Laing et al. 1994; Grimes et al. 2004) is that steep spectrum ($\alpha > 0.5$) 
HEGs and quasars belong to the same parent population, and that it is the orientation of the source 
in the sky that decides whether it will appear as an HEG or a quasar, the latter when the major radio-axis happens to be within a 
certain critical angle ($\xi_{c}$) around the observer's line of sight. HEGs and quasars, in all other respects,  
are considered to be intrinsically the same. In this orientation-based unified scheme (OUS), 
because of the smaller inclinations of the radio axes of the quasars with respect to the observer's line of sight, 
the observed radio sizes of the quasars will be foreshortened due to the geometry and should appear systematically 
smaller than those of the HEGs. It is a popular notion that $\xi_{\rm c} \sim 45^{\circ}$ and that in the 3CRR catalogue 
the observed sizes of quasars are accordingly about a factor of two smaller as compared to those of radio galaxies 
(Barthel 1989; Urry \& Padovani 1995; Peterson 1997). 

\section{Results}
\begin{figure}
\scalebox{0.35}{\includegraphics{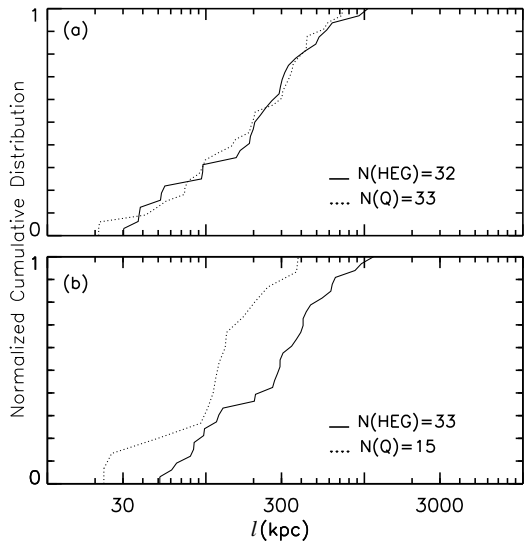}}
\caption{Normalized cumulative distributions of the linear size ($l$) of HEGs (continuous curves) and 
quasars (broken curves)  for the 3CRR sample (a) for region I (b) for region II. 
N(HEG) and N(Q) give the number of High Excitation galaxies and quasars respectively, in each case.}
\end{figure}

Recently it was shown that the relative size distributions of quasar and HEGs do not always show the projection effects,  
predicted by the OUS, when we compare the sources within different redshift bins (Singal 2014b). But what about when the 
size distributions are compared for different directions in the sky? To test this we divided the sample, starting from the first 
source in it, into two equal right ascension (RA) regions, region I from RA 0 to 12 and region II from 12 to 24 hours. 
While in region II the two size distributions differed by a factor of two or so, with quasar sizes being statistically smaller 
as one would expect due to the foreshortening in the OUS, in region I the sizes appeared statistically indistinguishable, contrary to the 
predictions of the OUS. This unexpected result prompted us to check their numbers as well in these two sky regions, since to be consistent 
with the OUS predictions, not only their relative sizes but their relative numbers should also 
differ by a factor of about two for a `canonical' value of $\xi_{\rm c} \sim 45^{\circ}$. And again we found that 
while in region II the number of quasars was indeed about half that of HEGs, but in region I there were as many quasars 
as the HEGs, contrary to what expected according to the OUS. 
Figure 1 shows normalized cumulative plots of the linear size distributions of HEGs and quasars in the two regions. 
In region II we do notice the quasar sizes (as well as numbers) to be smaller than those of the 
HEGs by a factor of about two. However in region I, the differences, if any, in radio sizes or numbers are hardly seen and 
a Kolmogorov-Smirnov test shows that 
the two distributions are statistically almost indistinguishable, thereby punching a hole in the unification scheme. 
Not only does this seem to be a very strong evidence against the OUS 
(after all the OUS could not hold good in just one half of the sky), but it seems that there could be much more at stake here than 
just the validity of the OUS.

In the OUS, the ratios in the sizes and numbers of HEGs and quasars could change with redshift depending upon details of the model used 
as, for example, in the receding-torus-type scheme (Lawrence 1991; Hill et al. 1996) where the critical   
angle ($\xi_{c}$) may be evolving with redshift or luminosity. But in any case the ratio should not vary with the direction in sky. 
Therefore while any variations in numbers or sizes with redshift one could try to put down to some sort of 
cosmological evolution of their properties, irrespective of whether or not unified scheme holds good, 
but the same type of escape route cannot be available for a variation (over and above what might be due to statistical fluctuations) 
in the sky distribution. Further, any effects of zone of avoidance ($\pm 10^\circ$ around the galactic plane, $b=0$) 
should proportionally be the same for both HEGs and quasars, without affecting their number ratios. 

It should be noted that even within the unification scheme, 
HEGs and quasars observationally are not identical and each class has distinct properties and they are identifiable or distinguishable 
as separate type of objects observationally and each of them should have their own isotropic distribution and
there should be no sky-position dependent effect between the two.  
In fact with or without the unified schemes, from the isotropy expected from the cosmological principle, the number of 
any type of distant extragalactic objects should not vary with direction in sky, apart from the statistical fluctuations. 
A close investigation showed that while the HEGs, which are the largest number of the 3CRR constituents, are quite uniformly 
distributed over the observed sky, the quasars are quite unevenly distributed. While about two thirds 
(33 out of a total of 48) quasars in the sample lie in region I, the remainder one third (15 out of 48) appear 
in region II. In a priori chosen division of the sky in two adjacent and contiguous regions, for a random distribution of 
the sources one expects to get a binomial distribution. The probability of such a deviation in a binomial distribution to occur  
at $(33-15)/\sqrt {48} \sim 2.6 \sigma$ level due to statistical fluctuations is only $\sim 0.01$ (Bevington \& Robinson 2003). 
\begin{figure}
\scalebox{0.4}{\includegraphics{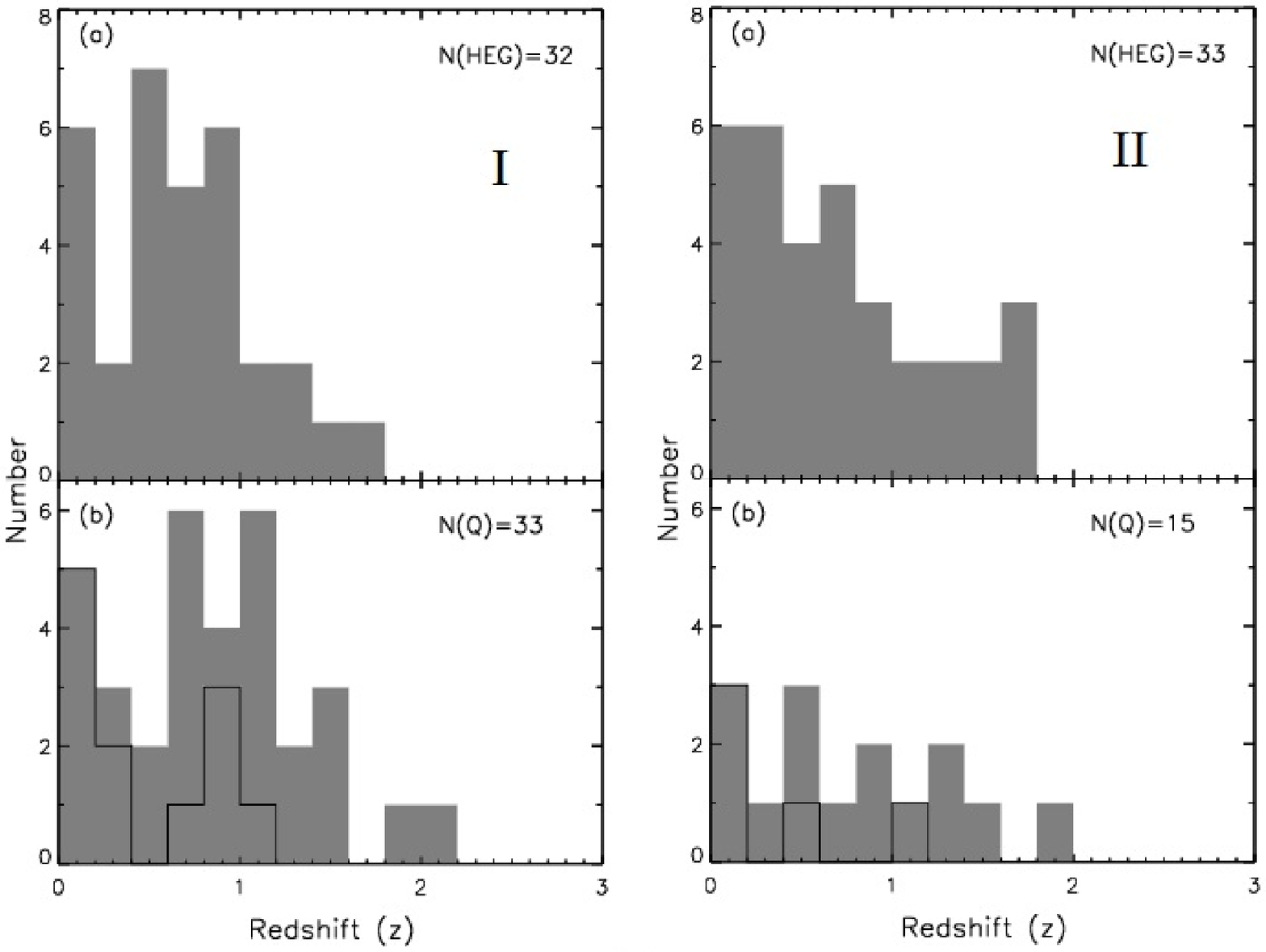}}
\caption{Histograms of the redshift distributions of the 3CRR sample in regions I and II of the sky  
(a) for HEGs (b) for quasars. In the lower panels, the regions under the overlaid darker lines represent WQs or BLRGs. 
N(HEG) and N(Q) give the number of high excitation galaxies and quasars respectively, in each region.}
\end{figure}

Could the anisotropy in the distribution of quasars have any local Supercluster or some other local origin? 
Such a thing, if any, should show up as a difference in the redshift distributions in the two regions. 
Figure 2 shows the redshift distributions of HEGs and quasars. Apart from the total number of quasars being less in region II, 
there does not appear to be any gross changes with redshift in the distribution of quasars and as well of HEGs  
in the two regions. This almost rules out the possibility that the quasar anisotropy has any local origin. 
Even the weak quasars (WQs), which like the other quasars are also proportionally less in region II, have redshift 
distributions which are very similar in the two regions, so any anomaly in quasar distribution is certainly not due to the 
presence of a differential number of WQs. 
\begin{table}[t]
\caption{Counts of radio sources in two regions of the sky.}
\begin{tabular}{@{\hspace{0mm}}c@{\hspace{3mm}}c@{\hspace{4mm}}c@{\hspace{4mm}}c@{\hspace{4mm}}c@{\hspace{4mm}}c@{\hspace{4mm}}c}
\hline
Sky region  & N(HEG) & N(Q)& N(LEG) & N(FR1)\\
\hline
I + II & 65 & 48 & 17 & 23  \\
I & 32 & 33  & 12 & 6  \\
II & 33 & 15  & 5 & 17 \\
\hline
\end{tabular}
\end{table}

Figure 3 shows a normalized cumulative plot of HEG and quasar distributions in RA in sky. The sky distributions of HEGs and 
quasars appear very different, with slightly more than two thirds of all quasars lying in region I, while HEGs are 
distributed quite evenly over the sky. Also plotted in the figure are distributions 
of LEGs and FR1s. These too show very uneven distributions, with about 70\% of LEGs lying in regions I and only about 1/4th of 
FR1s in that region. Overall the percentage of FR1s varies substantially between the two regions, 
while in region 1 there are only 7\% of the total sources as FR1s, in region II the percentage is as much as 24\%.
Table 1 gives the number counts of different type of sources in the two regions of the sky. 
The probabilities of such a deviation to occur in a binomial distribution 
at $11/\sqrt {23} \sim 2.3 \sigma$ level due to statistical 
fluctuations is $\sim 0.02$ for FR1s. Further, as asymmetries of quasars and FR1s would have independent binomial probabilities, 
if these were due to random statistical fluctuations, 
then their combined probability of occurrence due to being simply a statistical fluctuation is only about 
$0.01 \times 0.02 \sim 2 \times 10^{-4}$, i.e., a $4 \sigma$ result. 

Similarly LEGs also show an asymmetric distribution though at somewhat lower level (Table 1); while in 
region I there are 12 LEGs, in region II there are only 5 LEGs, implying a $7/\sqrt {17} \sim 1.7 \sigma$ deviation. 
If we include their probability of occurrence at $\sim 0.09$ as well, then the total combined probability becomes 
$\sim 2 \times 10^{-5}$. It should be noted that LEGs and FR1s, which may have overlap in some 
of their properties, are otherwise different type of objects classified by their distinct radio  
properties, e.g., their different radio morphologies. Moreover LEGs are of higher radio 
luminosities than the FR1s and are seen at relatively much higher redshifts. Therefore their uneven distributions 
are not a result of a mix-up in their classifications. Quasars of course stand apart, being the most energetic and most 
distant of these objects.
\begin{figure}[t]
\scalebox{0.45}{\includegraphics{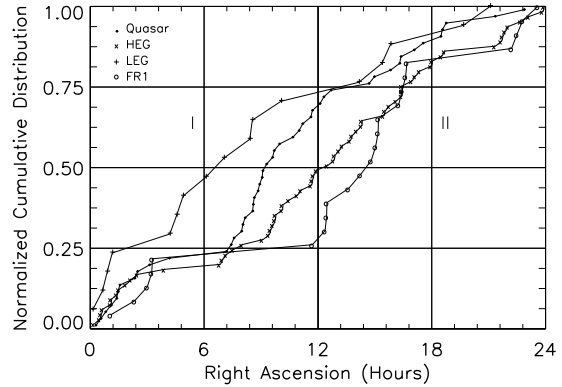}}
\caption{Normalized cumulative sky distributions of various objects in the 3CRR sample plotted against RA.}
\end{figure}

To ensure that there is nothing amiss in our probability calculations, we also did Monte Carlo simulations by throwing quasars,  
LEGS and FR1s randomly in sky and counting the number of times we get a distribution like that in Table 1. A total of 1000 simulations 
were done, every time starting with a different seed for a random number generator, while in each simulation 100,000 different random throws 
of 48 quasars, 17 LEGs and 23 FR1s were made. Thus in total a 100 million independent trials were made and out of these 1856 
cases were found to have deviations equal to those in Table 1, implying a probability consistent with our calculations of 
$\sim 2 \times 10^{-5}$. 
\begin{figure}[t]
\scalebox{0.45}{\includegraphics{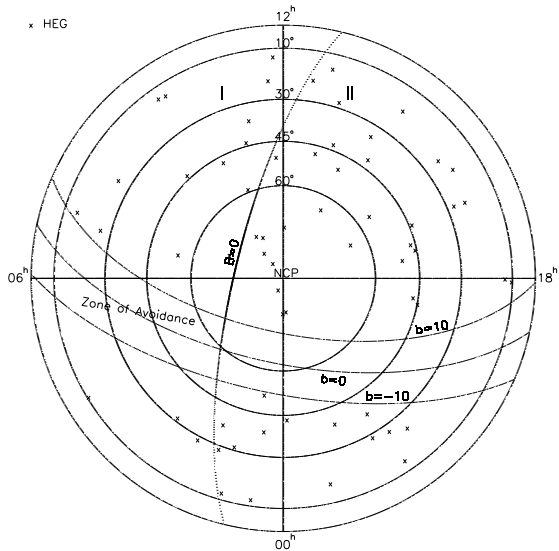}}
\caption{Sky distribution of HEGs from the 3CRR sample shown in an equal-area projection of the Northern hemisphere, 
centered on the North Celestial Pole. Region I extends in right ascension  from 0 to 12 hour and lies to the left of the vertical line passing through 
the NCP while region II lies to the right of it. The zone of avoidance is shown by a band of $\pm 10^\circ$ about the galactic plane ($b=0^\circ$).
Also shown is the Super-galactic plane ($B=0^\circ)$.}
\end{figure}
 
Figure 4 shows a Lambert azimuthal equal-area projection, mapping the Northern hemisphere onto a circular disc centered on NCP and  
accurately representing areas in all regions of the hemisphere. 
All points on a circle at a declination $\delta$ in sky are represented by a circle of radius $\propto \sqrt{1-\sin \delta}$ on the disc.
The figure shows plot of HEGs from the 3CRR catalogue; the distribution seems to be fairly uniform on the sky.
\begin{figure}[t]
\scalebox{0.45}{\includegraphics{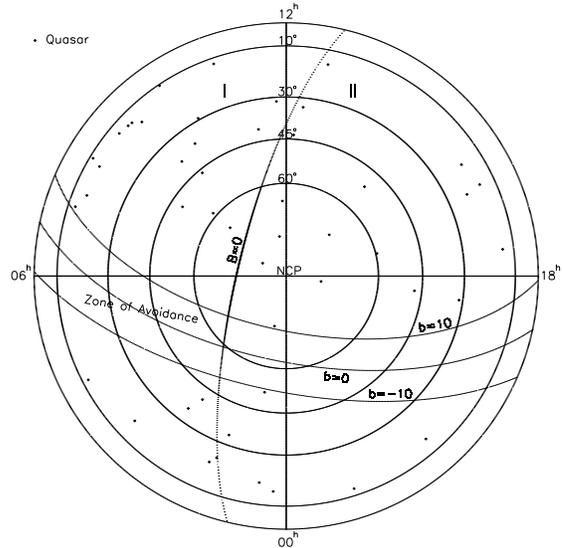}}
\caption{Sky distribution of quasars from the 3CRR sample shown in an equal-area projection of the Northern hemisphere,
centered on the North Celestial Pole. Region I extends in right ascension  from 0 to 12 hour and lies to the left of the vertical line passing through 
the NCP while region II lies to the right of it. The zone of avoidance is shown by a band of $\pm 10^\circ$ about the galactic plane ($b=0^\circ$).
Also shown is the Super-galactic plane ($B=0^\circ)$.}
\end{figure}

Figure 5 shows plot of quasars from the 3CRR catalogue on the sky. To a first order, this division of sky in regions I and II 
happens to yield almost the maximum  asymmetry visible in the quasar distribution, and it amounts to passing a great circle between 
the equinoxes (intersection points of the equatorial plane and the ecliptic) and the north celestial pole (NCP). 
First thing we want to be sure is that the observed number of quasars in region I being double or so of that in region II is not due to 
any instrumental/observational selection effects in these two regions in the 3CRR catalogue. This is guaranteed by the fact that 
virtually no difference is seen between the numbers of HEGs from these two regions in the same catalogue (Figure 4), which could not have happened 
if there were any such selection effects. It confirms that the quasar anomaly is not due to any observational selection effects, as any 
selection effects would not treat HEGs and quasars differentially, which were first radio selected and only later categorized as HEGs or 
quasars from their optical/infrared properties.  The same argument can also be applied for the absence of any influence
of our Galaxy on various distributions, as the Galaxy could not have affected distribution of different type of objects differently. Even 
otherwise the quasar asymmetry in Figure 5 seems to have no correlation with the galactic plane.

Comparing the regions between RA 06 to 12 hours and 12 to 18 hours in top half of Figure 5, we notice that there are 22 quasars between 
RA 06 to 12 hours while there are only 10 between 12 to 18 hours, giving a ratio of 2.2 in quasar numbers between these two regions. 
Actually with about 10\% of the region from 06 to 12 hours overlapping the zone of avoidance, one would rather expect a 
proportionally smaller number of quasars in that region as compared to that in RA 12 to 18 hours, contrary to what actually seen. 
The total number of sources in the bottom half of the figure is less as compared to that in the top half, mainly because of a large fraction 
of area in the bottom half overlapping with the zone of avoidance. But even there as well a ratio of 2.2 is found between the region 
0 to 6 hour (11 quasars) and that between 18 to 24 hour (5 quasars). From this it is clear that the asymmetry in quasar distribution  
is not due to a local excess (i.e., any local clustering) in neighborhood of some point in sky and that this excess in RA range 0 to 12 
hour as compared to 12 to 24 hour is fairly widely distributed. Also there seems to be no effect of the Galactic latitude on the quasar 
distribution outside the zone of avoidance.  The Super-galactic plane ($B=0$) too does not seem to 
have any relation with the distribution of quasars on sky. This of course is expected as quasars are at much higher redshifts as compared to 
that of the local Virgo supercluster. Therefore being two to three orders of magnitude more distant than the Virgo Supercluster, 
quasars can in no way be physically related to it or some other local objects. 
\begin{figure}[t]
\scalebox{0.45}{\includegraphics{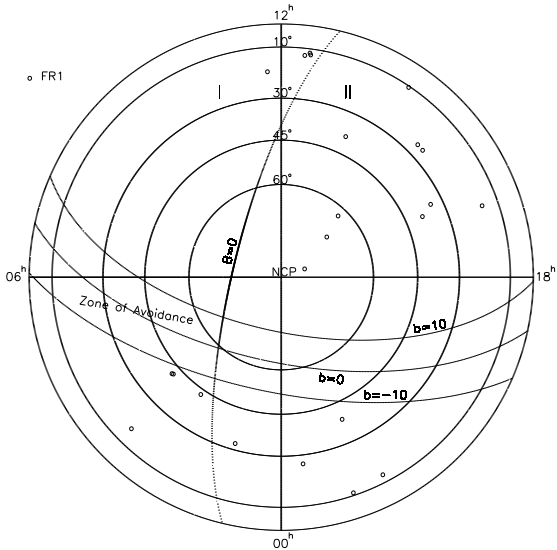}}
\caption{Sky distribution of FR1 type of objects from the 3CRR sample shown in an equal-area projection of the Northern hemisphere, 
centered on the North Celestial Pole. Region I extends in right ascension  from 0 to 12 hour and lies to the left of the vertical line passing through 
the NCP while region II lies to the right of it. The zone of avoidance is shown by a band of $\pm 10^\circ$ about the galactic plane ($b=0^\circ$).
Also shown is the Super-galactic plane ($B=0^\circ)$.}
\end{figure}

Figure 6 shows the distribution of FR1 types of radio galaxies in the sky. It is clear that FR1 radio galaxies 
also have a highly asymmetric number distribution between the two regions, 
though in opposite sense to that of quasars. The distribution is particularly asymmetric about the 
line joining the Autumn equinox (RA=12 hour) to the NCP. While there are 13 FR1s between RA=12 to 18 hour, there is only 1 FR1 radio 
galaxy between RA range from 6 to 12 hour, and that too lies close to the boundary at 12 hour. The area covered by galactic plane 
in the region RA 06 to 12 hours is only $\sim 10\%$, so that does not resolve the asymmetry. Nor is  
this order of magnitude difference explained even if we exclude a couple of FR1s (M84; M87 or Virgo A) 
which lie close to the Super-galactic plane ($B=0$). If we drop the two FR1s close to the Super-galactic plane and adjust for the $10\%$ 
galactic plane coverage, then we have approximately 10 versus 1 FR1s in the two regions which may imply a $9/\sqrt {11} \sim 2.7 \sigma$ 
fluctuation, with a $\sim 0.007$ chance probability.
\section{Discussion and conclusions}
It is interesting that while relatively low redshift (up to $z\sim 0.2$) FR1s have excess between 12 to 18 hr RA, 
high redshifted quasars (up to $z\sim 2$) 
have an excess in the RA range from 6 to 12 hour, in direction where FR1s are almost non-existent. This shows not only an 
anisotropic universe but also a direct evidence of the presence of large scale inhomogeneities. It should be noted that the scale 
spanned by FR1s in the 
universe (up to $\sim$ a gigaparsec) is almost an order of magnitude larger than the scale at which inhomogeneities 
(Super-clusters, Great-Wall, Voids etc.) have till now been seen through optical observations.
And of course quasars further cover a scale an order of magnitude larger than FR1s. This in fact is the largest scale in which discrete 
objects have been seen in the universe and any anisotropy or inhomogeneity on that scale is certainly a cause of worry as it will 
negate the cosmological principle. 

These results are robust. There is little likelihood that these anomalies could be due to some missing or even spurious sources in the 
3CRR catalogue, a radio complete sample of sources, in the sense that all source 
above the sensitivity limit of the catalogue have been detected and listed. 

It is to be noted that a large scale dipole anisotropy in radio source distribution at much fainter levels was seen earlier, 
and was interpreted due to motion of the solar system with respect to an average universe. The derived direction of motion 
matched with that inferred from the CMBR, though the magnitude was found to be about a factor of four larger (Singal 2011) 
than for CMBR (Lineweaver et al. 1996; Hinshaw et al. 2009). These apparently anomalous 
results have recently been vindicated by the findings of two independent groups (Rubart \& Schwarz 2013; Tiwari et al. 2014). 

However the anisotropies pointed out here in the 3CRR sample could not be caused by a motion of the solar system as it 
could not give rise to different anisotropies for different kind of objects. We have seen that while powerful HEGs numbers are evenly 
distributed, quasars and LEGs have more numbers in region I, but the less luminous FR1's are found to be more in region II. 
It is as if different regions of the sky were more amenable to one kind of source types than the other.
Nor could these be attributed to some effect of 
our Galaxy or some effect of local Supercluster. Any such things would have affected all type of different objects in 
roughly the same way, but as we have seen the HEGs, LEGs, quasars and FR1s have very different asymmetries in their distributions. 

There is certainly something intriguing. Is there a breakdown of the Copernican principle as things seen in two regions of sky, 
divided purely by a coordinate system based on earth's orientation in space, show very large anisotropies in extragalactic source 
distributions? Why should the equinox points should have any bearing on the large scale distribution of matter in the 
universe?  The only way to still retain the cosmological principle will be to doubt the reliability of the 3CRR survey, 
which will come very much of a surprise to almost all radio astronomers who take the 3CRR sample to be a true representation of strong 
radio source population. It should be noted that in the last three decades, since the 
3CRR sample was formed (Laing et al. 1983), there have been a few, if any, changes due to addition of missing sources 
or deletions of spurious sources, and it is unlikely that the problem would get resolved that way. 
Many more deeper surveys covering all sky are certainly required in order to resolve this enigma, but even if deeper or 
more complete southern surveys show the absence of these anisotropies in the sky distribution of quasars and/or other 
radio galaxies, it will still remain to be explained why these anomalies are present in the strong 3CRR sample in the Northern 
hemisphere. After all many important studies like the number counts, luminosity function and/or cosmological evolution of 
other properties of radio population have 
been made using the 3CRR source distributions as an important ingredient, where an implicit assumption was an isotropic distribution 
of radio sources in the 3CRR sample (or at least presence of no such large anomalies), 
whether for quasars or for other radio objects. Even if in future it does turn out that one could explain away these anomalies due to 
some ill-understood subtle local effect, it might still require at least a rethinking on some of these earlier results. 
The OUS at least seems to be ousted as it cannot be valid only in one half of the sky as implied by the number and size ratios.
A further confirmation of the asymmetries will of course be much vicissitudinous for all astronomers and cosmologists as well, 
since cosmological principle is the basis on which almost all modern cosmological theories depend upon as a starting point.
   
For the fore-mentioned apparent alignment in the CMBR in one particular direction through space, 
it has to be kept in mind that 
all such observations are obscured by the disc of the Milky Way galaxy, and one has to be extra careful 
while interpreting the data. Even there have been speculations whether solar system dust could give rise to sizable level of 
microwave emission or absorption, leading to a correlation with the ecliptic (Dikarev et al. 2009). But no such effect will be expected 
in the number distributions of discrete sources. 
The normals to the four quadrupole and octopole planes are aligned with the direction of the equinoxes and so does the dipole direction 
representing the overall motion of the solar system in the universe (Schwarz et al. 2004; 
Copi et al. 2010).  Also our plane dividing the two 
regions of asymmetry passes through the same two equinox points.
But it is not clear whether the asymmetries seen by us are related to that in the CMBR, 
as it is not presently possible to see if the 
anomalous distribution of radio sources is really related to ecliptic coordinates as the region covered by the 3CRR,  
unlike equatorial coordinates, is not divided equally in two ecliptic hemispheres. Perhaps an all-sky complete catalogue in future 
will help resolve this issue.
But irrespective of that there is no denying that from the large anisotropies present in the radio sky, independently 
seen both in the discrete source distributions and in the diffuse CMBR, the Copernican principle seems to be in jeopardy. 
\section*{Acknowledgements}
I thank Robert Antonucci for his comments, especially about the Copernican principle.

\end{document}